\documentclass[10pt]{article}
\usepackage{amssymb,amsmath,amsthm,fullpage,hyperref}

\newtheorem*{theorem}{Theorem}

\theoremstyle{remark}
\newtheorem{remark}{Remark}
\newtheorem{assumption}{Assumption}

\begin{document}

\begin{figure}
\begin{flushright}
SNU 08061
\end{flushright}
\end{figure}

\title{Continuous degeneracy of non-supersymmetric vacua}
\author{Zheng Sun\\
\normalsize\textit{Center for Theoretical Physics, Seoul National University, Seoul 151-747, Korea}\\
\normalsize\textit{E-mail:} \texttt{zsun@phya.snu.ac.kr}}
\date{}
\maketitle

\begin{abstract}
In global supersymmetric Wess-Zumino models with minimal K\"ahler potentials, F-type supersymmetry breaking always yields instability or continuous degeneracy of non-supersymmetric vacua. As a generalization of the original O'Raifeartaigh's result, the existence of instability or degeneracy is true to any higher order corrections at tree level for models even with non-renormalizable superpotentials. The degeneracy generically coincides the R-axion direction under some assumptions of R-charge assignment, but generally requires neither R-symmetries nor any assumption of generic superpotentials. The result also confirms the well-known fact that tree level supersymmetry breaking is a very rare occurrence in global supersymmetric theories with minimal K\"ahler potentials. The implication for effective field theory method in the landscape is discussed and we point out that choosing models with minimal K\"ahler potentials may result in unexpected answers to the vacuum statistics. Supergravity theories or theories with non-minimal K\"ahler potentials in general do not suffer from the existence of instability or degeneracy. But very strong gauge dynamics or small compactification dimension reduces the K\"ahler potential from non-minimal to minimal, and gravity decoupling limit reduces supergravity to global supersymmetry. Instability or degeneracy may appear in these limits. Away from these limits, a large number of non-SUSY vacua may still be found in an intermediate region.
\end{abstract}

\section{Introduction}

It is known ever since the discovery of spontaneous supersymmetry breaking \cite{O'Raifeartaigh:1975pr, Intriligator:2007cp} that there is a massless Goldstino in any SUSY breaking theory involving chiral fields, i.e.\ Wess-Zumino models \cite{Wess:1973kz}. One related result is that such a non-SUSY vacuum has continuous degeneracy, a.k.a.\ pseudomoduli space or flat direction. This result has been proven for any renormalizable Lagrangian \cite{O'Raifeartaigh:1975pr}. The result also shows that metastable SUSY breaking is a very rare occurrence in global SUSY theories with minimal K\"ahler potentials. In generic cases, non-SUSY extrema usually have mass matrices with negative eigenvalues which indicate tachyonic instability. Only with fine tuning of the potential does the non-SUSY vacuum has a vanishing mass matrix and remains stable up to quadratic. The degeneracy is true only at tree level. Loop corrections \cite{Coleman:1973jx} lift up the flat direction and generate a potential for the moduli. That is why the degeneracy is also called ``pseudomoduli'' space.

There is a connection between SUSY breaking and R-symmetries described by Nelson-Seiberg theorem \cite{Nelson:1993nf}. For a generic model without fine tuning, a $\operatorname{U}(1)$ R-symmetry for the superpotential is a necessary and sufficient condition for SUSY breaking. The R-symmetry needs to be broken to have non-zero Majorana gaugino masses. If the breaking is spontaneous, it implies the existence of a massless Goldstone boson, the R-axion, which coincides the degeneracy in many, but not all, O'Raifeartaigh's models considered to date. This statement is not true for some non-generic superpotential. There are exceptions where R-symmetries do not guarantee SUSY breaking, or SUSY is broken without any R-symmetry. Moreover, Nelson-Seiberg theorem only tells the non-existence of SUSY vacua, but does not guarantee the (meta)stability of any specific non-SUSY vacuum. As we show later in this paper, models with the coincidence of the R-axion and the degeneracy share some characteristic R-charge assignment. But the existence of continuous degeneracy is a more general result, requires neither the existence of R-symmetries nor the assumption of generic superpotentials.

Recent studies of effective field theory method in the string landscape \cite{Dine:2005yq, Giudice:2006sn} suggest that a large number of metastable non-SUSY vacua is only possible due to higher order corrections in the K\"ahler potential (or supergravity corrections, which we discuss later). In a simple example with a minimal K\"ahler potential and a non-renormalizable superpotential of a single chiral field \cite{Giudice:2006sn}, it is shown that any non-SUSY extremum is either a saddle point or has exact continuous degeneracy. The latter case requires a series of coefficients to be set to zero, which makes the occurrence of non-SUSY vacua very rare. The purpose of this paper is to investigate the general case with arbitrary number of chiral fields, i.e.:

\begin{theorem}
In any global supersymmetric Wess-Zumino model with a minimal K\"ahler potential, an arbitrary number of chiral fields and a superpotential which can be any renormalizable or non-renormalizable holomorphic function of chiral fields, F-type SUSY breaking always yields instability or exact continuous degeneracy of the non-SUSY vacuum at tree level.
\end{theorem}

A proof of the result has been provided in \cite{Ray:2006wk}. In this paper we demonstrate the proof in a different notation which gives insight to some related questions, e.g.\ the direction of the degeneracy and the relation to R-symmetries and the R-axion. We also emphasize the existence of instability if there is no degeneracy, which composes the base of the argument for the implication for the landscape. One should notice that simply integrating out extra fields and reducing to one field problem does not work, since this usually leads to a non-minimal K\"ahler potential which invalidates the proof. To make a rigorous proof, it is necessary to study the ``full'' theory with all chiral fields present.

Surely there are many situations which fall outside the scope of this theorem. The metastability is only a local result except along the degeneracy direction. So there may be some SUSY ``true'' vacuum separated from the non-SUSY one by a potential wall. The theorem can only be applied to F-type SUSY breaking in a low energy effective Wess-Zumino theory. SUSY breaking by Fayet-Iliopoulos terms \cite{Fayet:1974jb} is not discussed in this paper. There are a class of calculable theories where the scale of SUSY breaking can be made small compared to the scale at which the gauge dynamics become strong \cite{Affleck:1984xz}. One can integrate out the strongly coupled gauge fields and there will be only chiral fields left in the SUSY breaking sectors we are interested in. But the integrating out procedure usually produces non-minimal K\"ahler potentials, which invalidate the theorem except in some non-generic model where the K\"ahler metric is flat along the SUSY breaking field strength direction \cite{Aldrovandi:2008sc}. Also as one goes further to study supergravity theories, the theorem can not be applied. Although moduli spaces are common in supergravity theories, they are generally irrelevant to the question whether SUSY is broken or not. In the gravity decoupling limit where the Planck mass is much larger than other energy scales in the theory, the SUGRA scalar potential reduces to global SUSY one, and the theorem can be applied if one has a minimal K\"ahler potential. But again, in realistic models one usually gets non-minimal K\"ahler potentials. So in realistic model building there is little need to worry about the constraint of the theorem.

Because of these situations, this theorem has more methodological importance in simple effective models than realistic model building. In the study of the string landscape \cite{Douglas:2006es, Denef:2007pq}, effective field theory method is an useful tool to give argument about the vacuum statistics \cite{Dine:2005yq, Giudice:2006sn}. As a consequence of the theorem, one has to consider a non-minimal K\"ahler potential in the effective model in order to get a non-zero measure distribution of non-SUSY vacua. The K\"ahler potential may reduces to minimal in the limit where the SUSY breaking scale is extremely small compared to the scale of gauge dynamics or compactification, then the theorem may be applied. SUGRA theories with minimal K\"ahler potentials can also produce a large number of non-SUSY vacua for the landscape, but the gravity decoupling limit changes the result, as we have mentioned before. Away from these two limits, there is an intermediate region where one may find a large number of non-SUSY vacua which we are interested in.

The outline of the paper is as follows. In section 2 we present a proof for the simple case with only one field as a demonstration of the general proof algorithm. In section 3 we provide the proof for the general case of the theorem. In section 4 we give several remarks on the direction of degeneracy, applicability of the theorem and the question of stability. In section 5 we show that under some assumptions of R-charge assignment the R-axion can be extended to a flat complex plane and coincides the degeneracy from the theorem, and we also give out examples and exceptions to show that the existence of the degeneracy does not require R-symmetries or any assumption of genericity. In section 6 we discuss in the EFT method study of the landscape how the theorem affect the SUSY-breaking scale distribution of the vacua and different limits, and we give out suggestions on choosing the right EFT model for the study.

\section{Proof for the one field case}

Let us begin with only one chiral field. The proof for the general case with multiple chiral fields can be motivated by the one field case, but with more complexity and subtlety. It is helpful to use this simple case to show the similar outline of the proof.

\begin{proof}
Consider a global SUSY theory of one chiral field $z$. By field redefinition, we can set the field value to vanish at the vacuum:
\begin{equation}
z|_{\text{vacuum}} = 0 \ .
\end{equation}
Since the superpotential is a holomorphic function of $z$, we can expand it at the origin:
\begin{equation}
W = \sum_n a_n z^n \ .
\end{equation}
For simplicity, in this paper the sum of power index always goes from $0$ to $\infty$. Since we are relaxing the renormalizability condition, the expansion can have non-zero $a_n$ for any positive integer $n$. SUSY breaking can be checked by the field strength
\begin{equation}
\partial_z W = \sum_n (n+1)a_{n+1} z^n \ .
\end{equation}
Plugging it into the expression for the scalar potential, we have
\begin{equation}
\begin{split}
V& = \lvert \partial_z W \rvert^2 = \sum_{n,m} (n+1)(m+1)a_{n+1}^* a_{m+1} \bar{z}^n z^m\\
 & = a_1^* a_1 + (2a_1^* a_2 z + \text{c.c.}) + 4a_2^* a_2 \bar{z} z + (3a_1^* a_3 z^2 + \text{c.c.}) + O(z^3)
\end{split}
\end{equation}
where c.c.\ means complex conjugate. For convenience of the following proof, the expansion up to quadratic is shown above.

If SUSY is unbroken,
\begin{equation}
\partial_z W|_{z=0} = 0 \quad \Rightarrow \quad a_1 = 0 \ .
\end{equation}
The mass matrix is always positive definite:
\begin{equation}
V = \lvert 2 a_2 z \rvert^2 + O(z^3) \ ,
\end{equation}
which is consistent with the fact that any SUSY extremum is also a true minimum.

If SUSY is broken,
\begin{equation}
\partial_z W|_{z=0} \ne 0 \quad \Rightarrow \quad a_1 \ne 0 \ .
\end{equation}
The condition for extremum is
\begin{equation}
\partial_z V|_{z=0} = 0 \quad \Rightarrow \quad a_2 = 0 \ .
\end{equation}
Expanding the potential at the origin, we have
\begin{equation}
\begin{split}
V& = a_1^* a_1 + (3a_1^* a_3 z^2 + \text{c.c.}) + O(z^3)\\
 & \le a_1^* a_1 \quad \text{for} \quad z = \epsilon e^{i(\pi - \arg(a_1^* a_3))/2}, \quad \epsilon > 0 \ .
\end{split}
\end{equation}
Thus the extremum has quadratic instability unless the equality is satisfied, which implies
\begin{equation}
a_3 = 0 \ .
\end{equation}
But now we have a zero mass matrix. One has to check if there is cubic or higher order instability.

We continue the proof by mathematical induction. Suppose we have checked the theory up to $(K-1)$th order, and found the non-SUSY extremum at the origin is either unstable or has to satisfy
\begin{equation}
a_1 \ne 0, \quad a_2 = \ldots = a_K = 0, \quad K>1 \ .
\end{equation}
We have already checked above that for $K=3$ this statement is true\footnote{The induction can actually start from $K=2$. Similarly in the proof for the general case of the next section the induction can also start one step earlier.}. Now expanding the potential at the origin, we have
\begin{equation}
\begin{split}
V& = a_1^* a_1 + ((K+1)a_1^* a_{K+1} z^K + \text{c.c.}) + O(z^{K+1})\\
 & \le a_1^* a_1 \quad \text{for} \quad z = \epsilon e^{i(\pi - \arg(a_1^* a_{K+1}))/K}, \quad \epsilon > 0 \ .
\end{split}
\end{equation}
Thus the extremum has $K$th order instability unless the equality is satisfied, which implies
\begin{equation}
a_{K+1} = 0 \ .
\end{equation}
So the statement is also true up to $K$th order. By the axiom of induction, we have the conclusion that the the non-SUSY extremum is either unstable or has to satisfy
\begin{equation}
a_1 \ne 0, \quad a_n = 0, \quad n>1 \ .
\end{equation}
Thus the $V$ is independent of $z$, the vacuum has exact degeneracy along the direction $z$ to all higher orders.
\end{proof}

The proof provided above is to demonstrate the outline of the proof for the general case. There is a more simple proof for the one field case, as shown in follows\footnote{We thank Satoshi Yamaguchi for pointing this out to us.}. The non-SUSY extremum has
\begin{equation}
V|_{\text{vacuum}} = \lvert \partial_z W \rvert^2 > 0 \ .
\end{equation}
So we can take the logarithm of $V$ in the neighborhood of the extremum:
\begin{equation}
\log V = 2 \operatorname{Re}(\log \partial_z W) \ .
\end{equation}
Using the theorem in complex analysis that the real part of a holomorphic function can not have a minimum in a finite region, we conclude that the non-SUSY extremum can not be a minimum of $\log V$, thus can not be a minimum of $V$. The only way it could be stable is to have a constant $V$, or degeneracy along the direction $z$. But this simple proof can not be easily generalized to multiple fields.

\section{Proof for the general case}

We will give the general proof with similar algorithm of the one field case: Expand the superpotential at the extremum, and use mathematical induction to prove that a set of coefficients vanish. To show the degeneracy we only need to work out the expansion along some specific direction. This algorithm is also similar to the one used in \cite{Ray:2006wk}.

\begin{proof}
Consider a global SUSY theory of $d$ chiral fields. By field redefinition, we can set the field values to vanish at the vacuum:
\begin{equation}
z_i|_{\text{vacuum}} = 0, \quad i=1, \ldots, d \ .
\end{equation}
Since the superpotential is a holomorphic function of all $z_i$, we can expand it at the origin:
\begin{equation}
W = \sum_{n_i} a_{n_1 \ldots n_d} z_1^{n_1} \ldots z_d^{n_d} \ .
\end{equation}
SUSY breaking can be checked by the field strength
\begin{equation}
\partial_j W = \sum_{n_i} (n_j + 1)a_{n_1 \ldots n_j +1 \ldots n_d}  z_1^{n_1} \ldots z_d^{n_d} \ .
\end{equation}
Plugging it into the expression for the scalar potential, we have
\begin{equation}
\begin{split}
V =& \sum_j \lvert \partial_j W \rvert^2 = \sum_j \sum_{n_i, m_i} (n_j + 1)(m_j + 1)a_{n_1 \ldots n_j +1 \ldots n_d}^* a_{m_1 \ldots m_j + 1 \ldots m_d} \bar{z_1}^{n_1} \ldots \bar{z_d}^{n_d} z_1^{m_1} \ldots z_d^{m_d}\\
  =& \sum_i a_{(i,1)}^* a_{(i,1)} + (\sum_i 2a_{(i,1)}^* a_{(i,2)} z_i + \sum_{i \ne j} a_{(i,1)}^* a_{(i,1)(j,1)} z_j + \text{c.c.}) +\\
   & + \sum_i 4a_{(i,2)}^* a_{(i,2)} \bar{z_i} z_i + (\sum_{i \ne j} 2a_{(i,2)}^* a_{(i,1)(j,1)} \bar{z_i} z_j + \text{c.c}) + \sum_{i \ne j, i \ne k} a_{(i,1)(j,1)}^* a_{(i,1)(k,1)} \bar{z_j} z_k +\\
   & + (\sum_i 3a_{(i,1)}^* a_{(i,3)} z_i^2 + \sum_{i \ne j} 2a_{(i,1)}^* a_{(i,1)(j,2)} z_j^2 + \sum_{i \ne j} 2a_{(i,1)}^* a_{(i,2)(j,1)} z_i z_j +\\
   & \phantom{+(} + \sum_{i \ne j, i \ne k, j \ne k} a_{(i,1)}^* a_{(i,1)(j,1)(k,1)} z_j z_k + \text{c.c.}) +\\
   & + O(z_i^3) \ .
\end{split}
\end{equation}
Same as we have done in the one field case, the expansion up to quadratic is shown above. The sum of field index goes from $1$ to $d$. The notation for the index of $a$ is defined as
\begin{equation}
a_{n_1 \ldots n_d} = a_{(1,n_1) \ldots (d,n_d)}
\end{equation}
and $n_i = 0$ part is omitted in the newly defined index, e.g.\ $a_{310 \ldots 0} = a_{(1,3)(2,1)}$.

If SUSY is unbroken,
\begin{equation}
\partial_j W|_{z_i = 0} = 0 \quad \Rightarrow \quad a_{(j,1)} = 0 \ .
\end{equation}
The mass matrix is always positive definite:
\begin{equation}
V = \lVert \sum_j A_{ij} z_j \rVert^2 + O(z_i^3),
\quad A_{ij} =
\begin{cases}
2a_{(i,2)},& i=j\\
a_{(i,1)(j,1)},& i \ne j
\end{cases} \ ,
\end{equation}
which is consistent with the fact that any SUSY extremum is also a minimum.

If SUSY is broken, suppose we have some non-zero field strength in terms of the basis $z'_i$:
\begin{equation}
\partial'_i W = F'_i \ .
\end{equation}
One can always find a unitary rotation such that
\begin{equation}
\sum_j U_{ij} F'_j = (\lVert F \rVert, 0, \ldots, 0), \quad U \in \operatorname{SU}(d) \ .
\end{equation}
Then the field redefinition
\begin{equation} \label{eq:3-10}
z'_j = \sum_i U_{ij} z_i \quad \text{or} \quad z_j = \sum_i U^\dagger_{ij} z'_i
\end{equation}
makes the field strength have only one non-zero component:
\begin{equation}
\partial_i W = \sum_j \frac{\partial z'_j}{\partial z_i} \partial'_j W = \sum_j U_{ij} \partial'_j W = (\lVert F \rVert, 0, \ldots, 0) \ .
\end{equation}
Notice the K\"ahler potential is still minimal after this field redefinition. So we have
\begin{equation} \label{eq:3-20}
\begin{gathered}
\partial_1 W|_{z_i = 0} \ne 0 \quad \Rightarrow \quad a_{(1,1)} \ne 0 \ , \\
\partial_j W|_{z_i = 0} = 0 \quad \Rightarrow \quad a_{(j,1)} = 0, \quad j>1 \ .
\end{gathered}
\end{equation}
The condition for extremum is
\begin{equation}
\partial_j V|_{z_i = 0} = 0 \quad \Rightarrow \quad a_{(1,2)} = a_{(1,1)(j,1)} = 0, \quad j>1 \ .
\end{equation}

Expanding the potential at the origin along the direction $(z_1, 0, \ldots, 0)$, we have
\begin{equation}
\begin{split}
V(z_1, 0, \ldots, 0)& = a_{(1,1)}^* a_{(1,1)} + (3a_{(1,1)}^* a_{(1,3)} z_1^2 + \text{c.c.}) + O(z_1^3)\\
 & \le a_{(1,1)}^* a_{(1,1)} \quad \text{for} \quad z_1 = \epsilon e^{i(\pi - \arg(a_{(1,1)}^* a_{(1,3)}))/2}, \quad \epsilon > 0 \ .
\end{split}
\end{equation}
Thus the extremum has quadratic instability unless the equality is satisfied, which implies
\begin{equation} \label{eq:3-30}
a_{(1,3)} = 0 \ .
\end{equation}
In the case where \eqref{eq:3-30} is satisfied, the expansion becomes
\begin{equation}
\begin{split}
V(z_1, 0, \ldots, 0)& = a_{(1,1)}^* a_{(1,1)} + (4a_{(1,1)}^* a_{(1,4)} z_1^3 + \text{c.c.}) + O(z_1^4)\\
 & \le a_{(1,1)}^* a_{(1,1)} \quad \text{for} \quad z_1 = \epsilon e^{i(\pi - \arg(a_{(1,1)}^* a_{(1,4)}))/3}, \quad \epsilon > 0 \ .
\end{split}
\end{equation}
Thus the extremum has cubic instability unless the equality is satisfied, which implies:
\begin{equation} \label{eq:3-40}
a_{(1,4)} = 0 \ .
\end{equation}
Now expanding the potential along the direction $(z_1, 0, \ldots, z_i, \ldots, 0), \ i>1, \ z_1 \gg z_i \gg z_1^2$ in the case where \eqref{eq:3-30}\eqref{eq:3-40} is satisfied, we have
\begin{equation}
\begin{split}
V(z_1, 0, \ldots, z_i, \ldots, 0)& = a_{(1,1)}^* a_{(1,1)} + (2a_{(1,1)}^* a_{(1,2)(i,1)} z_1 z_i + \text{c.c.}) + O(z_i^2)\\
 & \le a_{(1,1)}^* a_{(1,1)} \quad \text{for} \quad z_1 z_i = \epsilon e^{i(\pi - \arg(a_{(1,1)}^* a_{(1,2)(i,1)}))}, \quad \epsilon > 0 \ .
\end{split}
\end{equation}
Notice the magnitude of $z_i$ is important for the ordering of the expansion\footnote{The next term of the expansion is of order $z_i^2$ or $z_1^4$, both are higher than order $z_1 z_i$ under the condition. Also we have $z_i^2 \gg z_1^4$.}. The extremum has quadratic instability unless the equality is satisfied, which implies
\begin{equation}
a_{(1,2)(i,1)} = 0, \quad i>1 \ .
\end{equation}

Suppose we have checked the theory up to order $z_1^{2K-1}$ and $z_1^{K-1} z_i$, and found the non-SUSY extremum at the origin is either unstable or has to satisfy
\begin{equation}
a_{(1,1)} \ne 0, \quad a_{(1,2)} = \ldots = a_{(1,2K)} = a_{(1,1)(i,1)} = \ldots = a_{(1,K)(i,1)} = 0, \quad i>1, \quad K>0 \ .
\end{equation}
We have already checked above that for $K=2$ this statement is true. Now expanding the potential at the origin along the direction $(z_1, 0, \ldots, 0)$, we have:
\begin{equation} \label{eq:3-50}
\begin{split}
V(z_1, 0, \ldots, 0)& = a_{(1,1)}^* a_{(1,1)} + ((2K+1)a_{(1,1)}^* a_{(1,2K+1)} z_1^{2K} + \text{c.c.}) + O(z_1^{2K+1})\\
 & \le a_{(1,1)}^* a_{(1,1)} \quad \text{for} \quad z_1 = \epsilon e^{i(\pi - \arg(a_{(1,1)}^* a_{(1,2K+1)}))/(2K)}, \quad \epsilon > 0 \ .
\end{split}
\end{equation}
Thus the extremum has $(2K)$th order instability unless the equality is satisfied, which implies
\begin{equation} \label{eq:3-60}
a_{(1,2K+1)} = 0 \ .
\end{equation}
In the case where \eqref{eq:3-60} is satisfied, the expansion becomes\footnote{The vanishing of $a_{(1,1)(i,1)}, \ldots, a_{(1,K)(i,1)}$ is important for this step since it suppresses the non-holomorphic terms up to order $\bar{z_1}^{K+1} z_1^K$.}
\begin{equation}
\begin{split}
V(z_1, 0, \ldots, 0)& = a_{(1,1)}^* a_{(1,1)} + ((2K+2)a_{(1,1)}^* a_{(1,2K+2)} z_1^{2K+1} + \text{c.c.}) + O(z_1^{2K+2})\\
 & \le a_{(1,1)}^* a_{(1,1)} \quad \text{for} \quad z_1 = \epsilon e^{i(\pi - \arg(a_{(1,1)}^* a_{(1,2K+2)}))/(2K+1)}, \quad \epsilon > 0 \ .
\end{split}
\end{equation}
Thus the extremum has $(2K+1)$th order instability unless the equality is satisfied, which implies
\begin{equation} \label{eq:3-70}
a_{(1,2K+2)} = 0 \ .
\end{equation}
Now expanding the potential along the direction $(z_1, 0, \ldots, z_i, \ldots, 0), \ i>1, \ z_1^K \gg z_i \gg z_1^{K+1}$ in the case where \eqref{eq:3-60}\eqref{eq:3-70} is satisfied, we have
\begin{equation}
\begin{split}
V(z_1, 0, \ldots, z_i, \ldots, 0)& = a_{(1,1)}^* a_{(1,1)} + ((K+1)a_{(1,1)}^* a_{(1,K+1)(i,1)} z_1^K z_i + \text{c.c.}) + O(z_i^2)\\
 & \le a_{(1,1)}^* a_{(1,1)} \quad \text{for} \quad z_1^K z_i = \epsilon e^{i(\pi - \arg(a_{(1,1)}^* a_{(1,K+1)(i,1)}))}, \quad \epsilon > 0 \ .
\end{split}
\end{equation}
Notice the magnitude of $z_i$ is important for the ordering of the expansion\footnote{The next term of the expansion is of order $z_i^2$ or $z_1^{2K+2}$, both are higher than order $z_1^K z_i$ under the condition. Also we have $z_i^2 \gg z_1^{2K+2}$.}. The extremum has $(K+1)$th order instability unless the equality is satisfied, which implies
\begin{equation}
a_{(1,K+1)(i,1)} = 0, \quad i>1 \ .
\end{equation}
So the statement is also true up to order $z_1^{2K+1}$ and $z_1^K z_i$. By the axiom of induction, we have the conclusion that the the non-SUSY extremum is either unstable or has to satisfy
\begin{equation} \label{eq:3-80}
a_{(1,1)} \ne 0, \quad a_{(1,n)} = a_{(1,n)(i,1)} = 0, \quad i>1, \quad n>0 \ .
\end{equation}
Thus the vacuum has exact degeneracy along the direction $(z_1, 0, \ldots, 0)$ to all higher orders.
\end{proof}

\section{Several remarks}

\begin{remark} \label{rm:4-10}
One can always choose a field basis so that the only field which breaks SUSY is the pseudomodulus which labels the complex plane of the degeneracy.
\end{remark}

This is clearly shown in the proof. The degeneracy is along $(z_1, 0, \ldots, 0)$ after the field redefinition \eqref{eq:3-10}. And the field strength \eqref{eq:3-20} has the form $\partial_i W = (\lVert F \rVert, 0, \ldots, 0)$. Notice this means the vector of the field strength always lives in the complex plane of the degeneracy.

\begin{remark} \label{rm:4-20}
The non-SUSY extremum can not be a maximum without continuous degeneracy.
\end{remark}

If the vacuum is unstable, one may ask if it can be a local maximum. From the proof we can see this is not possible if there is no degeneracy. Consider, for example, the extremum has $(2K)$th order instability along $(z_1, 0, \ldots, 0)$. Then we see in \eqref{eq:3-50} that
\begin{equation}
\begin{gathered}
a_{(1,2K+1)} \ne 0 \quad \Rightarrow \\
V(z_1, 0, \ldots, 0) > a_{(1,1)}^* a_{(1,1)}  \quad \text{for} \quad z_1 = \epsilon e^{-i \arg(a_{(1,1)}^* a_{(1,2K+1)})/(2K)}, \quad \epsilon > 0 \ .
\end{gathered}
\end{equation}
The result will be similar if the extremum has $(2K+1)$th order instability along $(z_1, 0, \ldots, 0)$. So we see the extremum can only be a saddle point if it does not have continuous degeneracy. On the other hand, it is possible to have a maximum with continuous degeneracy, e.g.\ in the model of \cite{Ellis:1982vi}, as we show at the end of this section.

\begin{remark}
The theorem can be applied to singular superpotentials, as long as the non-SUSY vacuum is not located at the singularity.
\end{remark}

A remormalizable superpotential $W$ can only have terms up to cubic. By integrating out heavy chiral fields or strongly coupled gauge fields \cite{Affleck:1984xz}, one can have a non-remormalizable but still holomorphic $W$. $W$ can have some singularity. But the vacuum which we are interested in must not be located at any singularity point, otherwise some of the ``heavy'' fields are actually massless, thus can not be integrated out. So there is a finite neighborhood of the non-SUSY vacuum where the $W$ can be expanded, and we can prove as we showed before that there is continuous degeneracy in this neighborhood. The degeneracy can be extended along the pseudomodulus direction all through the field space. This can be shown by apagogical argument. Suppose the degeneracy is only in a finite region. Then we can choose a point on the boundary of this region, and expand $W$ at this point. The expansion is flat along the degeneracy direction, and valid in a finite neighborhood of the point, which contradicts the assumption that the point is on the boundary of the degeneracy region. So the exact continuous degeneracy always exists, as long as the non-SUSY vacuum is not located at the singularity, which is a consequence of the properly done integrating out.

\begin{remark}
The result is the same if there are other fields which do not enter the SUSY breaking sector.
\end{remark}
Other fields may include gauge fields such that the chiral fields in the SUSY breaking sector are neutral under the gauge symmetry, and standard model fields or other visible sector which the SUSY breaking is mediated with. Such fields do not change the SUSY breaking mechanism, so the theorem still works.

\begin{remark}
The theorem only gives a necessary but not sufficient condition for SUSY breaking. Moreover, the metastable non-SUSY vacuum with degeneracy, if existing, is just a local minimum. There may be some SUSY vacuum or runaway direction in the theory which the non-SUSY vacuum may decay into.
\end{remark}

During the proof, only the metastable condition along a certain direction is checked for the extremum. So the extremum may have tachyonic or higher order instability along other directions. Even if the vacuum is metastable, there may still be some SUSY vacuum separated from the non-SUSY one by a potential wall. Then the non-SUSY vacuum may decay into the SUSY one which is the ``true'' minimum by quantum tunneling \cite{Coleman:1977py, Coleman:1980aw}. To show this, one example is given in \cite{Ellis:1982vi}. Suppose we have a superpotential
\begin{equation}
W = \lambda z_1 z_3 (z_3 - m) + \mu z_2 (z_3 - m) \ ,
\end{equation}
and taking all the coefficients to be real and non-zero. The field strength is
\begin{equation}
\partial_1 W = \lambda z_3 (z_3 - m), \quad \partial_2 W = \mu (z_3 - m), \quad \partial_3 W = \lambda z_1 (2z_3 - m) + \mu z_2 \ .
\end{equation}
There is a SUSY minimum at $z_3 = m$, a non-SUSY local minimum at $z_3 \approx \mu^2 / (\lambda^2 m)$ if $\mu^2 \ll \lambda^2 m^2$ and a local maximum at $z_3 \approx m/2 - \mu^2 / (\lambda^2 m)$ which is between the two minima\footnote{Although two of the extrema given here are just approximate solutions, all of them definitely exist. One can find exact solutions by solving cubic equations which set the derivatives of the scalar potential to zero.}. All the three extremum have continuous degeneracy along $z_2 = - \lambda z_1 (2z_3 - m) / \mu$ with their respective values of $z_3$. So we see the possibility to have a metastable non-SUSY vacuum which may decay into a SUSY one, and a maximum with degeneracy which is mentioned above in the comment of remark \ref{rm:4-20}. It is also possible that the non-SUSY vacuum may decay to a runaway direction, which is quite common in SUSY breaking models with R-symmetries \cite{Ferretti:2007ec, Carpenter:2008wi}.

\section{Relation to R-symmetries}

According to Nelson-Seiberg theorem \cite{Nelson:1993nf}, in global SUSY theories, generically SUSY is broken if and only if there is a $\operatorname{U}(1)$ R-symmetry for the superpotential. The R-symmetry needs to be broken to have non-zero Majorana gaugino masses. If it is spontaneously broken, there is a one-dimensional degeneracy of the non-SUSY vacuum associated with the Goldstone boson, the R-axion. Here we will see, under some assumptions, the R-axion can be extended to a whole flat complex plane at tree level which coincides the continuous degeneracy from the theorem. The assumptions are:

\begin{assumption} \label{as:5-10}
The K\"ahler potential is minimal.
\end{assumption}

\begin{assumption} \label{as:5-20}
The field which breaks the SUSY has R-charge $2$.
\end{assumption}

\begin{assumption} \label{as:5-30}
Other fields either are R-neutral or have zero vacuum expectation values.
\end{assumption}

Although these assumptions sound very non-generic, as partly pointed out by \cite{Shih:2007av}, they are satisfied by many of the O'Raifeartaigh models considered to date, such as the original O'Raifeartaigh model which we will show later in this section. Assumption \ref{as:5-10} ensures the existence of the degeneracy from the theorem, because Nelson-Seiberg theorem actually allows non-minimal (and even R-asymmetric) K\"ahler potentials which invalidates our theorem. Assumption \ref{as:5-30} implies the R-symmetry does not change vacuum expectation values of fields except the R-charge $2$ field mentioned in assumption \ref{as:5-20}. So the R-axion is only associated to that R-charge $2$ field. And according to remark \ref{rm:4-10} of the theorem, the whole R-charge $2$ field plane is a pseudomoduli space. So we see the R-axion lives in the complex plane of the degeneracy from our theorem. When this coincidence happens, the R-axion must have R-charge $2$ as mentioned in assumption \ref{as:5-20}. This can be seen from the expansion of the superpotential $W$. According to \eqref{eq:3-80}, $W$ has a non-vanishing term $a_{(1,1)} z_1$. Since $W$ has R-charge $2$, the SUSY breaking field $z_1$ which is associated to the R-axion also has R-charge $2$.

Another way to see the R-axion must have R-charge 2 to get the flat plane is to consider the complexification of the symmetry group as described in \cite{Carpenter:2008wi}. The R-symmetry rotates fields as well as the superpotential, and also the SUSY breaking field strength:
\begin{equation}
\partial_i W \rightarrow e^{i (2-q_i) \alpha} \partial_i W, \quad \alpha \in \mathbb{R} \ .
\end{equation}
The manifold of SUSY solutions is larger than what the R-symmetry implies, because one can take the parameter of the R-symmetry $\alpha$ to be complex and the zero-valued field strength solution is still unaffected by the complexified symmetry. For a non-SUSY vacuum, zero-valued components of the field strength remain invariant. But for non-zero $\partial_i W$, it transforms under the R-symmetry and its magnitude varies if $\alpha$ is taken to be complex unless $q_i = 2$. So if and only if the assumption \ref{as:5-20} is satisfied, the scalar potential is invariant under the complexified R-symmetry, and the R-axion can be extended to a whole flat complex plane. Notice to get this result we only used assumption \ref{as:5-20}. Assumption \ref{as:5-30} makes sure that this complex plane coincides with the degeneracy from the theorem. If it is not satisfied, one just gets a plane of degeneracy different than the one from our theorem.

We are to demonstrate these relations by examples. First we start with an O'Raifeartaigh model proposed in many literatures and textbooks, e.g. \cite{Dine:2007zp}. The model has three chiral fields and the superpotential
\begin{equation} \label{eq:5-30}
W = \lambda z_1 (z_3^2 - m^2) + \mu z_2 z_3
\end{equation}
where $z_1, z_2$ have R-charge $2$ and $z_3$ has R-charge $0$. The components of the SUSY breaking field strength
\begin{equation}
\partial_1 W = \lambda (z_3^2 - m^2), \quad \partial_2 W = \mu z_3, \quad \partial_3 W = 2\lambda z_1 z_3 + \mu z_2
\end{equation}
can not be set to zero simultaneously, so there is no SUSY vacuum for this model. We need to minimize the potential
\begin{equation}
V = \lvert \lambda \rvert^2 \lvert z_3^2 - m^2 \rvert^2 + \lvert \mu \rvert^2 \lvert z_3 \rvert^2 + \lvert 2\lambda z_1 z_3 + \mu z_2 \rvert^2 \ .
\end{equation}
Assuming all the coefficients are real, positive\footnote{One can make non-zero complex coefficients real and positive by field redefinition by phases. So making such assumption does not lose genericness. Same argument applies to the coefficients of other models in this section.} and satisfy
\begin{equation} \label{eq:5-40}
\mu^2 > 2\lambda^2 m^2 \ ,
\end{equation}
the non-SUSY vacuum is
\begin{equation} \label{eq:5-50}
z_2 = z_3 = 0, \quad z_1 = \text{arbitrary value} \ .
\end{equation}
We see there is continuous degeneracy along $z_1$. It corresponds to the R-axion direction since the R-symmetry rotates $z_1$. Also according to remark \ref{rm:4-10}, it is the same degeneracy as the the one appears in the theorem because the only non-zero field strength component is $\partial_1 W$. $z_1$ has R-charge $2$, and other fields have zero vacuum expectation values. So this model satisfies all assumptions which we have made, and consequently we have the coincidence. 

If instead of \eqref{eq:5-40}, the coefficients are real, positive and satisfy
\begin{equation}
\mu^2 < 2\lambda^2 m^2 \ ,
\end{equation}
then $z_3$ will get non-zero vacuum expectation value. The vacuum is
\begin{equation}
z_3 = r = \pm \sqrt{m^2 - \frac{\mu^2}{2\lambda^2}}, \quad 2\lambda r z_1 + \mu z_2 = 0, \quad \mu z_1 - 2\lambda r z_2 = \text{arbitrary value} \ .
\end{equation}
We redefine the fields as we have done in the proof:
\begin{equation}
z'_1 = A^{-1} (\mu z_1 - 2\lambda r z_2), \quad z'_2 = A^{-1} (2\lambda r z_1 + \mu z_2), \quad z'_3 = z_3, \quad A = \sqrt{4\lambda^2 r^2 + \mu^2} \ .
\end{equation}
The normalization factor $A$ is used to make the redefinition a unitary transformation so that the K\"ahler potential remains minimal. $z'_1, z'_2$ still have R-charge $2$, and $z'_3$ has R-charge $0$. One can check that the only non-zero field strength component is $\partial'_1 W$. Although $z'_3$ has non-zero vacuum expectation value, it is R-neutral. So all assumptions are satisfied and we still have the coincidence.

When such coincidence happens, one needs to consider loop corrections to determine whether the R-symmetry is broken or not. The above example has only R-charge $2$ and $0$ fields. According to \cite{Shih:2007av}, the one-loop Coleman-Weinberg potential stabilizes the pseudomodulus at the origin thus the R-symmetry is preserved by the vacuum. There is an example in \cite{Shih:2007av} with different R-charge assignment which spontaneously breaks the R-symmetry at loop level. It also satisfies all assumptions here. So the same coincidence happens at tree level.

The assumptions which we propose give a necessary (but not sufficient) condition for the coincidence of the R-axion extension and the degeneracy. To find an exception one needs to dissatisfy at least one of them. One such example appears in \cite{Carpenter:2008wi}. It has some trace of the original O'Raifeartaigh model \eqref{eq:5-30} but has been modified a lot. The model has five chiral fields and the superpotential
\begin{equation}
W = \lambda z_1 (z_4 z_5 - m^2) + \mu z_2 z_4 + \nu z_3 z_5 + \tau z_4^3
\end{equation}
where $z_1, \ldots, z_5$ have R-charge $2$, $4/3$, $8/3$, $2/3$, $-2/3$ respectively. The components of the SUSY breaking field strength
\begin{equation}
\begin{gathered}
\partial_1 W = \lambda (z_4 z_5 - m^2), \quad \partial_2 W = \mu z_4, \quad \partial_3 W = \nu z_5, \\
\partial_4 W = \lambda z_1 z_5 + \mu z_2 + 3 \tau z_4^2, \quad \partial_5 W = \lambda z_1 z_4 + \nu z_3
\end{gathered}
\end{equation}
can not be set to zero simultaneously, so there is no SUSY vacuum for this model. We need to minimize the potential
\begin{equation}
V = \lvert \lambda \rvert^2 \lvert z_4 z_5 - m^2 \rvert^2 + \lvert \mu \rvert^2 \lvert z_4 \rvert^2 + \lvert \nu \rvert^2 \lvert z_5 \rvert^2 + \lvert \lambda z_1 z_5 + \mu z_2 + 3 \tau z_4^2 \rvert^2 + \lvert \lambda z_1 z_4 + \nu z_3 \rvert^2 \ .
\end{equation}
Assuming all the coefficients are real, positive and satisfy
\begin{equation}
\mu \nu < \lambda^2 m^2 \ ,
\end{equation}
the non-SUSY vacuum satisfies
\begin{equation}
\lvert \mu z_4 \rvert = \lvert \nu z_5 \rvert, \quad z_4 z_5 = m^2 - \frac{\mu \nu}{\lambda^2}, \quad \lambda z_1 z_5 + \mu z_2 + 3 \tau z_4^2 = 0, \quad \lambda z_1 z_4 + \nu z_3 = 0 \ .
\end{equation}
There is also another extremum of $V$ where $z_2 = \ldots = z_5 = 0$ and $z_1$ labels the degeneracy, but it has higher $V$. The solution we provide above is actually the global minimum of the potential. The degeneracy from the theorem is along the direction
\begin{equation}
z'_1 = A^{-1} (\mu \nu z_1 - \lambda \nu z_2 z_5 - \lambda \mu z_3 z_4), \quad A = \sqrt{\mu^2 \nu^2 + \lambda^2 \mu^2 \lvert z_4 \rvert^2 + \lambda^2 \nu^2 \lvert z_5 \rvert^2} \ ,
\end{equation}
which should be viewed as a linear redefinition of $z_1, z_2, z_3$ with $z_4, z_5$ fixed. So $z'_1$ which breaks SUSY is a mixture of R-charge $2$, $4/3$, $8/3$ fields. $z_4, z_5$ have non-zero vacuum expectation values
\begin{equation}
z_4 = \nu r e^{i \theta}, \quad z_5 = \mu r e^{-i \theta}, \quad \theta \in \mathbb{R}, \quad r = \sqrt{\frac{m^2}{\mu \nu} - \frac{1}{\lambda^2}} \ .
\end{equation}
The R-symmetry is spontaneously broken by the non-zero $r$ and the R-axion is labeled by $\theta$. So the total pseudomoduli space is of real dimension $3$: $2$ from our theorem and $1$ from the R-axion. This model does not satisfy both assumptions \ref{as:5-20} and \ref{as:5-30}. So it is not a surprise that the coincidence does not happen. Also notice the R-symmetry is already broken by vacuum expectation values of $z_4, z_5$ at tree level, rather than by the pseudomoduli at loop level as discussed in \cite{Shih:2007av}. Such tree level R-symmetry breaking models are investigated in \cite{Sun:2008va}.

Finally, one notices that Nelson-Seiberg theorem applies only to generic models. There are also non-generic SUSY breaking examples which possess no R-symmetry. Since the proof of the theorem does not need any assumption of genericness, the continuous degeneracy always exists no matter whether there is an R-symmetry or not. We can show this by modifying the original O'Raifeartaigh model \eqref{eq:5-30}, adding to the superpotential a term quadratic in $z_3$:
\begin{equation}
W = \lambda z_1 (z_3^2 - m^2) + \mu z_2 z_3 + M z_3^2 \ .
\end{equation}
The SUSY breaking field strength becomes
\begin{equation}
\partial_1 W = \lambda (z_3^2 - m^2), \quad \partial_2 W = \mu z_3, \quad \partial_3 W = 2\lambda z_1 z_3 + \mu z_2 + 2 M z_3 \ .
\end{equation}
This model possesses no R-symmetry. There is no consistent way to assign R-charges. Assuming all the coefficients are real, positive and satisfy the same condition as \eqref{eq:5-40}, the model has the non-SUSY vacuum given by the same field values of \eqref{eq:5-50}. There is still continuous degeneracy along $z_1$ even if there is no R-symmetry. This model is non-generic because small perturbation to $W$, such as $\epsilon z_2^2$, which is forbidden by the R-symmetry in the R-symmetric model but allowed here, will restore the SUSY vacuum at
\begin{equation}
z_1 = \frac{\mu^2}{4\lambda \epsilon} - \frac{M}{\lambda}, \quad z_2 = \pm \frac{\mu m}{2 \epsilon}, \quad z_3 = \mp m \ .
\end{equation}
There is no non-SUSY vacuum after this perturbation. Notice such a perturbation term is forbidden by the R-symmetry in the R-symmetric model but allowed here. So we see our theorem can be applied to more general models no matter whether the model is generic or not.

\section{Implication for the landscape}

The landscape \cite{Douglas:2006es, Denef:2007pq} was proposed to ``solve'' the cosmological constant problem. The concept was motivated partly by the recent study of flux compactifications in string theory \cite{Grana:2005jc} which suggests a huge number of metastable vacua, with different cosmological constants and other low-energy phenomenological properties. Statistical analysis \cite{Kumar:2006tn} has been used to study possible predictions of physical quantities such as the SUSY breaking scale, the gauge group and matter content, Yukawa couplings, and so on. The vacuum distribution can be computed from microscopic string theory considerations such as flux compactification superpotentials. To do these computation requires quite a lot of work. In many cases, effective field theory method \cite{Dine:2005yq, Giudice:2006sn} with simple assumption of parameter distribution, which is often known to be uniform from microscopic theories \cite{Denef:2004ze, Denef:2004cf}, is useful to demonstrate the same good results as more complicated microscopic theories can achieve. But because of our theorem, one should be careful not to choose a too simple EFT to study, otherwise one may get an unexpected result of the vacuum statistics which gives little possibility for SUSY breaking.

As already noticed in recent works \cite{Giudice:2006sn}, a relatively large number of metastable non-SUSY vacua in a global SUSY Wess-Zumino theory is only possible due to non-minimal corrections to the K\"ahler potential. Our theorem gives a good explanation for this phenomenon. It is seen in the proof that to get a metastable non-SUSY vacuum (with continuous degeneracy), a series of coefficients have to be set to zero, otherwise the vacuum has tachyonic or higher order instability. This means the distribution of metastable non-SUSY vacua only consists a hypersurface in the parameter space. It is a zero measure subset of the whole vacuum distribution if we assume that all the coefficients have a non-singular distribution. So non-SUSY vacua are extremely hard to find in the landscape of global SUSY models with minimal K\"ahler potentials. Limits of models which reduce to global SUSY with minimal K\"ahler potentials also suffer from the rareness of non-SUSY vacua, as we show later.

To overcome the constraint of the theorem, one way is to include higher order corrections in the K\"ahler potential, such as shown in the following example:
\begin{equation}
K = \bar{z} z - \frac{1}{4\Lambda^2} (\bar{z} z)^2, \quad W = M^2 z \ .
\end{equation}
The non-minimal part of K\"ahler potential gives mass to the field:
\begin{equation}
V = (1 - \frac{1}{\Lambda^2} \bar{z} z)^{-1} M^4 = M^4 + \frac{M^4}{\Lambda^2} \bar{z} z + O(z^4) \ .
\end{equation}
So the theorem does not hold for non-minimal K\"ahler potentials. The mass scale $M$ in the coefficient of $W$ is related to the SUSY breaking scale, and $\Lambda$ in the higher order correction of $K$ is related to, depending on the theory, either the scale at which the gauge dynamics becomes strong, or the compactification scale. In both cases one may take the limit that the scale $\Lambda$ is much larger than the SUSY breaking scale, then the K\"ahler potential reduces to minimal, and the mass term in $V$ vanishes. For general cases, because the coefficient of any higher order correction to $K$ always has negative mass dimension, we always have such a limit:
\begin{equation}
\lim_{\Lambda \rightarrow \infty} K = \sum_i h_{ij} \bar{z}_i z_j
\end{equation}
where $h_{ij}$ is a positive-definite Hermitian matrix. $h$ can be diagonalized by a unitary transformation $U^\dagger h U = \operatorname{diag}(h_1, \ldots, h_d), \ U \in \operatorname{SU}(d)$. Then by rescaling the fields it becomes an identity matrix. So the limit of the K\"ahler potential can take a minimal form:
\begin{equation}
\lim_{\Lambda \rightarrow \infty} K = \sum_i \bar{z}_i z_i \ .
\end{equation}
Then the theorem can be applied in the limit.

Another way to invalidate the theorem is to invoke local symmetries, a.k.a.\ supergravity. A SUGRA EFT model, even with a minimal K\"ahler potential, shows a similar distribution of the SUSY breaking scale as the model with a non-minimum K\"ahler potential \cite{Dine:2005yq}. But there is a small parameter in SUGRA models: the Planck length $1/M_P$. When $M_P$ is much larger than any other energy scales in the theory, the gravity decouples from other parts of the theory, and the SUGRA model is separated to a pure SUGRA part and a global SUSY part. The scalar potential for SUGRA, with the Planck scale written explicitly, has the form
\begin{equation}
V = e^{K / M_P^2} (K^{\bar{i} j} (\partial_i W + W \partial_i K / M_P^2)^* (\partial_j W + W \partial_j K / M_P^2) - 3 W^* W / M_p^2) \ .
\end{equation}
Taking the gravity decoupling limit $M_P \rightarrow \infty$, it reduces to the global SUSY scalar potential:
\begin{equation}
\lim_{M_P \rightarrow \infty} V = K^{\bar{i} j} (\partial_j W)^* \partial_i W \ .
\end{equation}
If the K\"ahler potential is minimal, i.e.\ $K^{\bar{i} j} = \delta^{ij}$, the theorem then can be applied, and one still rarely gets non-SUSY vacua.

Although in the above two limits non-SUSY vacua are still rarely found, there is a region where the SUSY breaking scale is not extremely small compared to the scale of gauge dynamics, compactification or the Planck mass. A large number of non-SUSY vacua may be find in this intermediate region. Notice these results are only true for tree level SUSY breaking. There are other branches of the landscape where SUSY is broken at loop level or non-perturbatively where the theorem can not be applied \cite{Dine:2005yq, Denef:2004ze, DeWolfe:2004ns, DeWolfe:2005gy, Dine:2004is, Dine:2005iw, Dine:2005gz}. Since realistic models usually have complicated K\"ahler potentials, which are obtained from either compactification of the microscopic theory or integrating out heavy fields, there is little to worry about in realistic model building.

To summarize: Global SUSY Wess-Zumino models with minimal K\"ahler potentials rarely have (tree level) metastable non-SUSY vacua. A relatively large number of metastable non-SUSY vacua can occur in SUGRA models with minimal K\"ahler potentials, but the occurrence becomes rare at the gravity decoupling limit. SUSY or SUGRA models with non-minimal K\"ahler potentials are not constrained by the theorem, so can have a non-zero measure distribution of SUSY breaking vacua. Exceptions happen at the limit where the mass scales in higher order coefficients of $K$ goes very large and $K$ reduces to minimal. When using EFT method to study the vacuum statistics of the landscape, one has to be careful whether it is proper to simplify the K\"ahler potential to minimal form.

\section*{Acknowledgement}

The author would like to thank Michael Dine, Michael R. Douglas, Soo-Jong Rey, Takao Suyama, Marcus A.C.Torres, Satoshi Yamaguchi and Hossein Yavartanoo for helpful discussions. Special thanks would like to be made to Michael Dine and Satoshi Yamaguchi for numerous suggestions for the early draft of this paper. This work is supported by BK-21 Program, KRF-2005-084-C00003, EU FP6 Marie Curie Research \& Training Networks MRTN-CT-2004-512194 and HPRN-CT-2006-035863 through MOST/KICOS.

\end{document}